\documentclass{appolb}
\usepackage{graphicx}
\usepackage[export]{adjustbox}
\usepackage{caption}
\usepackage{subcaption}
\usepackage{slashed}
\usepackage{epstopdf}
\begin{document}
\title{A covariant constituent-quark formalism for mesons%
\thanks{Presented at EEF70, Workshop on Unquenched Hadron Spectroscopy:\\
Non-Perturbative Models and Methods of QCD vs. Experiment}%
}
\author{S. Leit\~ao$^{1,2}$, A. Stadler$^{1,2}$,  M. T. Pe\~na$^{2,3}$,  E. P. Biernat$^2$
\address{$^1$Departamento de F\'isica, Universidade de \'Evora, 7000-671 \'Evora, Portugal\\
$^2$Centro de F\'isica Te\'orica de Part\'iculas (CFTP), Instituto Superior T\'ecnico, Universidade de Lisboa, 1049-001 Lisboa, Portugal\\
$^3$Departamento de F\'isica, Instituto Superior T\'ecnico (IST), Universidade de Lisboa, 1049-001 Lisboa, Portugal
}
}
\maketitle
\begin{abstract}
Using the framework of the Covariant Spectator Theory (CST) \cite{CST1} we are developing a covariant model formulated in Minkowski space to study mesonic structure and spectra. Treating mesons as effective $q\bar{q}$ states, we focused in \cite{linear} on the nonrelativistic bound-state problem in momentum space with a linear confining potential. Although integrable, this kernel has singularities which are difficult to handle numerically. In \cite{linear} we reformulate it into a form in which all singularities are explicitely removed. The resulting equations are then easier to solve and yield accurate and stable solutions. In the present work, the same method is applied to the relativistic case, improving upon the results of the one-channel spectator equation (1CSE) given in \cite{uzzo}.
\end{abstract}
  
\section{Introduction}
Due to the complex nature of hadronic matter it is challenging to find a comprehensive description of how quarks and gluons combine to form hadrons. In particular with the upcoming intense experimental activity dedicated to search for new exotic states, a better understanding of the conventional $q\bar{q}$ mesons is needed.
\\ 
\indent Phenomenological models that establish a link between lattice QCD and experimental data are important because they could help to reveal the connection between the hadronic mass spectrum and the underlying quark-gluon dynamics. 
\\ 
\indent In continuation of previous work by Gross, Milana and Savkli \cite{GM1,SG} using CST, we are developing a formalism that has potential to properly describe both light and heavy mesons, in a unified way. Moreover, the model is self-consistent because the quark self-energy is calculated from the same kernel that describes the quark-antiquark interaction. Recently \cite{QM,PFF}, the dressed quark mass function and pion electromagnetic form factor in impulse approximation have been calculated using this model and it has been shown that the model is consistent with the requirements of chiral symmetry \cite{pi-pi}.\\
\indent This work is the first application of the results of \cite{linear} to the 1CSE.  In \cite{linear} we address the problem of solving the momentum-space CST equations with the linear interaction in its nonrelativistic limit, in which the 1CSE becomes the Schr\"odinger equation, and develop a method to explicitely remove all the singularities of its kernel.
\section{CST bound-state equation}
\begin{figure}[tb]
\centering
\begin{minipage}[b]{1.00\linewidth}
\centering
\includegraphics[width=1.0\linewidth]{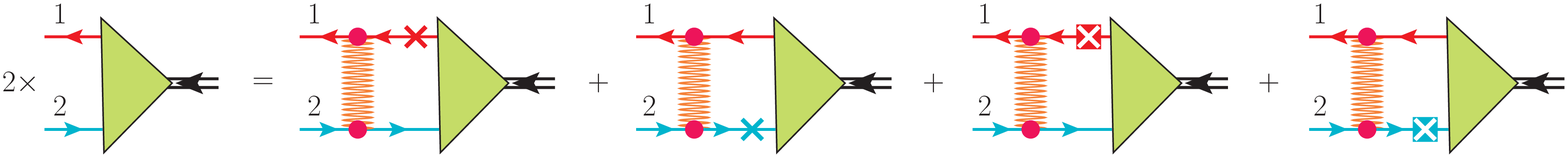}
\caption{\small Contributing diagrams for the 4CSE. A cross on a quark line indicates that only the positive-energy pole contribution of the corresponding propagator is kept in the loop integration, a cross inside a square refers to the respective negative-energy pole.}
\label{CST-BSE.eps}
\end{minipage}
\quad
\begin{minipage}[b]{0.63\linewidth}
\centering
\includegraphics[width=0.70\linewidth]{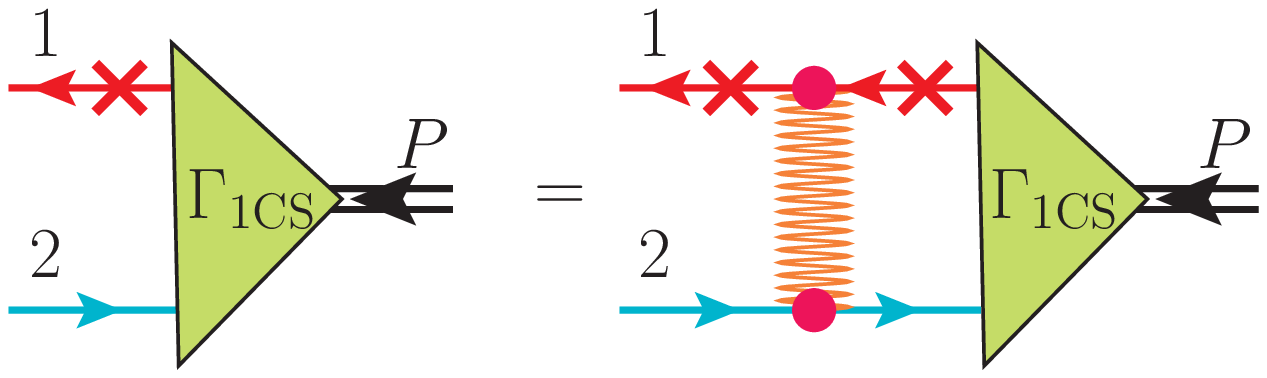}
\caption{\small Diagrammatic representation of the 1CSE.}
\label{1CS.eps}
\end{minipage}
\end{figure}

The CST bound-state equation emerges when we approximate the full Bethe-Salpeter (BS) equation for the vertex function $\Gamma$,
\begin{equation}
\Gamma_{BS}(p,P)=i\int \frac{d^4 k}{(2\pi)^4}\mathcal{V}(p,k;P)S_1(k_1)\Gamma_{BS}(k,P)S_2(k_2),
\label{Bs}
\end{equation}
 with total momentum $P$ and relative external and internal momentum $p$ and $k$, respectively. $S_i(k_i)=(m_{0i}-\slashed{k}_i+\Sigma_i(\slashed{k}_i) -i\epsilon)^{-1}$ ($i=1,2$) is the dressed propagator and $\Sigma_i(\slashed{k}_i)$ is the self-energy of quark $i$. This approximation consists of keeping only the pole contributions from the propagators at $k_{i0}=\pm E_{k_i}=\pm (m_i^2+\mathbf{k}^2)^{1/2}$, when the integration over $k_0$ is performed. If we symmetrize the contributions from both complex half-planes, we obtain an equation that is a three-dimensional reduction of Eq.~(\ref{Bs}) and has four contributing diagrams, depicted in Fig.~\ref{CST-BSE.eps}, each arising from  placing one particle on its positive/negative energy mass-shell.
When external legs are systematically placed on-shell in the diagrams of  Fig.~\ref{CST-BSE.eps}, a closed set of coupled equations emerges, the four-channel spectator equation (4CSE). However, to study heavy-light quark systems, with a large bound-state mass, it is sufficient to consider only the positive-energy pole contribution from the heavier particle 1, for $m_1>m_2$. The resulting equation, the 1CSE, is represented in Fig.~\ref{1CS.eps} and can be written as
\begin{equation}
\Gamma_{1CS}(p,P)=-\int \frac{d^3 k}{2E_{k_1}(2\pi)^3}V(p,k;P)\mathcal{O}^i_1\Lambda_1(\hat{k}_1)\Gamma_{1CS}(k,P)S_2(k_2)\mathcal{O}_2^i,
\label{gamma1CS}
\end{equation}
where $\mathcal{O}_1^i$ and $\mathcal{O}_2^i$ are Dirac matrices of type $i$ (scalar, vector, pseudoscalar), $ V(p,k;P)$ is the momentum-dependent part of the interaction, $\Lambda_1$ is the positive-energy projector and $\hat{k}_1=(E_{k_1},\mathbf{k})$ is the on-shell momentum of particle 1.
\section{Linear confining kernel}
In this preliminary work we are interested just in the linear confining part of the potential, 
\begin{equation}
\widetilde{V}(r)=\sigma r.
\end{equation}
 A covariant, relativistic generalization of this potential is
\begin{equation}
 V_L(p,k)=V_A(p,k)-2E_{p_1}(2\pi)^3\delta(\mathbf{p}-\mathbf{k})\int \frac{d^3 k'}{(2\pi)^3 2E_{k'_1}}V_A(p,k'),
\label{VL}
\end{equation}
with $V_A(p,k)\equiv-\displaystyle\frac{8\pi\sigma}{(p-k)^4}$.  In \cite{linear} we show that although this kernel is singular when $\mathbf{k}=\mathbf{p}$, when applied to any function of three-momenta, $F(\mathbf{p},\mathbf{k})$, one ends up with a Cauchy principal value integral ("$\mathrm{P}\!\!\int$"),
\begin{equation}
\small
\int \frac{d^3 k}{(2\pi)^3 2E_{k_1}}V_L(p,k)F(\mathbf{p},\mathbf{k})=\mathrm{P}\!\!\int \frac{d^3 k}{(2\pi)^3 2E_{k_1}}V_A(p,k)
\left[F(\mathbf{p},\mathbf{k})-F(\mathbf{p},\mathbf{p})\right].
\label{pot}
\end{equation}
We also show in \cite{linear} that the integrand of Eq.~(\ref{pot}) can be rearranged by means of a subtraction into two parts, one that is no longer singular and another one that contains a principal value singularity that can be integrated analytically. Besides treating the singularity, we also perform a partial-wave decomposition of Eq.~(\ref{gamma1CS}) and derive a singularity-free equation for an arbitrary partial wave $\ell$. Our results extend those of \cite{uzzo} in which only the $S$-wave case was computed.

\section{Numerical results}
 In order to solve Eq.~(\ref{gamma1CS}), we use a helicity representation (with helicity $\lambda=\pm 1/2$) for the 1CSE and expand it in terms of $\rho$ spinors $u^\rho_i(\mathbf{p},\lambda)$ ($\rho=\pm$) defined consistently with \cite{uzzo}. The most general form of the 1CSE for arbitrary interaction vertices and interaction kernel is
\begin{eqnarray}
\lefteqn{(E_{p_1}-\rho'E_{p_2})\Psi_{1\rho'}(p)-\sum_\rho\int \frac{d^3 k V(p,k)}{(2\pi)^34 E_{k_1}E_{k_2}}  } & &
  \nonumber \\
& & \hspace{-2mm}\times
\left[(2\lambda)^{\delta_{\rho'+}}\sum_{\lambda_1}\Theta^{++}_{1,\lambda\lambda_1}(p,k)(2\lambda_1)^{\delta_{\rho+}}
\Theta^{\rho\rho'}_{2,\lambda_1\lambda}(k,p)\right]
\!\!\Psi_{1\rho}(k)=\mu \Psi_{1\rho'}(p),
\label{matrix}
\end{eqnarray}
where $\Psi_{1\rho}(p)\equiv \displaystyle\frac{\rho \Gamma^\rho (p)}{E_{p_2}-\rho p_{20}}$,
and $\Gamma^\rho$ is obtained from the contraction of
\begin{equation}
\Gamma(p)=\Gamma_1(p)\gamma^5 + \Gamma_2(p)\gamma^5(m_2-\slashed{p}_2)
\label{gamma}
\end{equation}
 with the $\rho$ spinors. In the 1CSE, (\ref{gamma}) is the most general form of the vertex function for a pseudoscalar particle.\\
\indent Finally, the matrix elements of Eq.~(\ref{matrix}),
\begin{equation}
\Theta_{i,\lambda \lambda'}^{\rho\rho'}(p,k)\equiv \bar{u}_i^\rho(\mathbf{p},\lambda)\mathcal{O} u_i^{\rho'}(\mathbf{k},\lambda'),
\end{equation}
depend on the interaction kernel $\mathcal{O}^{(s)}=\mathbf{1}$ or $\mathcal{O}^{(v0)}=\gamma^0$ chosen. In this work we use a mixed kernel where one parameter $y$ dials continously between scalar and vector interaction, while preserving the nonrelativistic limit, which only differs in a sign  for both interactions:
\begin{equation}
\mathcal{V}(p,k)=V(p,k)\left[ (1-y) \mathbf{1}_1 \otimes \mathbf{1}_2  -y \gamma^0_1 \otimes \gamma^0_2 \right].
\end{equation}
These procedures enable us to transform the 1CSE  into an eigenvalue problem for the bound-state masses $\mu$ and the corresponding  wave functions, which are expanded in a basis of cubic B-splines as in \cite{uzzo}.
\begin{figure}[ht]
\captionsetup{justification=centering}
\begin{minipage}[b]{0.44\linewidth}
\centering
\includegraphics[width=1.07\linewidth]{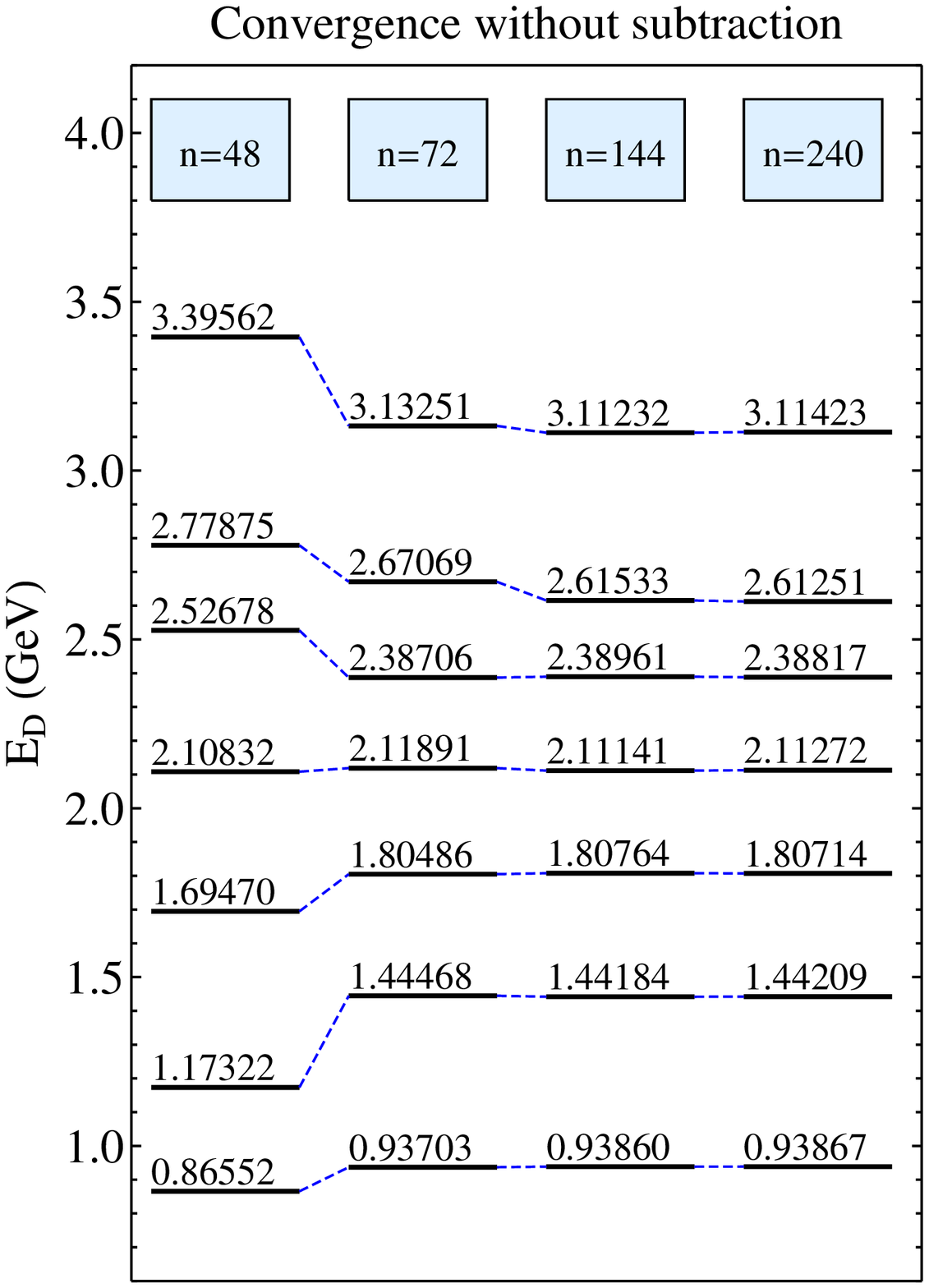}
\caption{\small Singular 1CSE results.}
\label{without}
\end{minipage}
\quad
\begin{minipage}[b]{0.485\linewidth}
\includegraphics[width=1.05\linewidth]{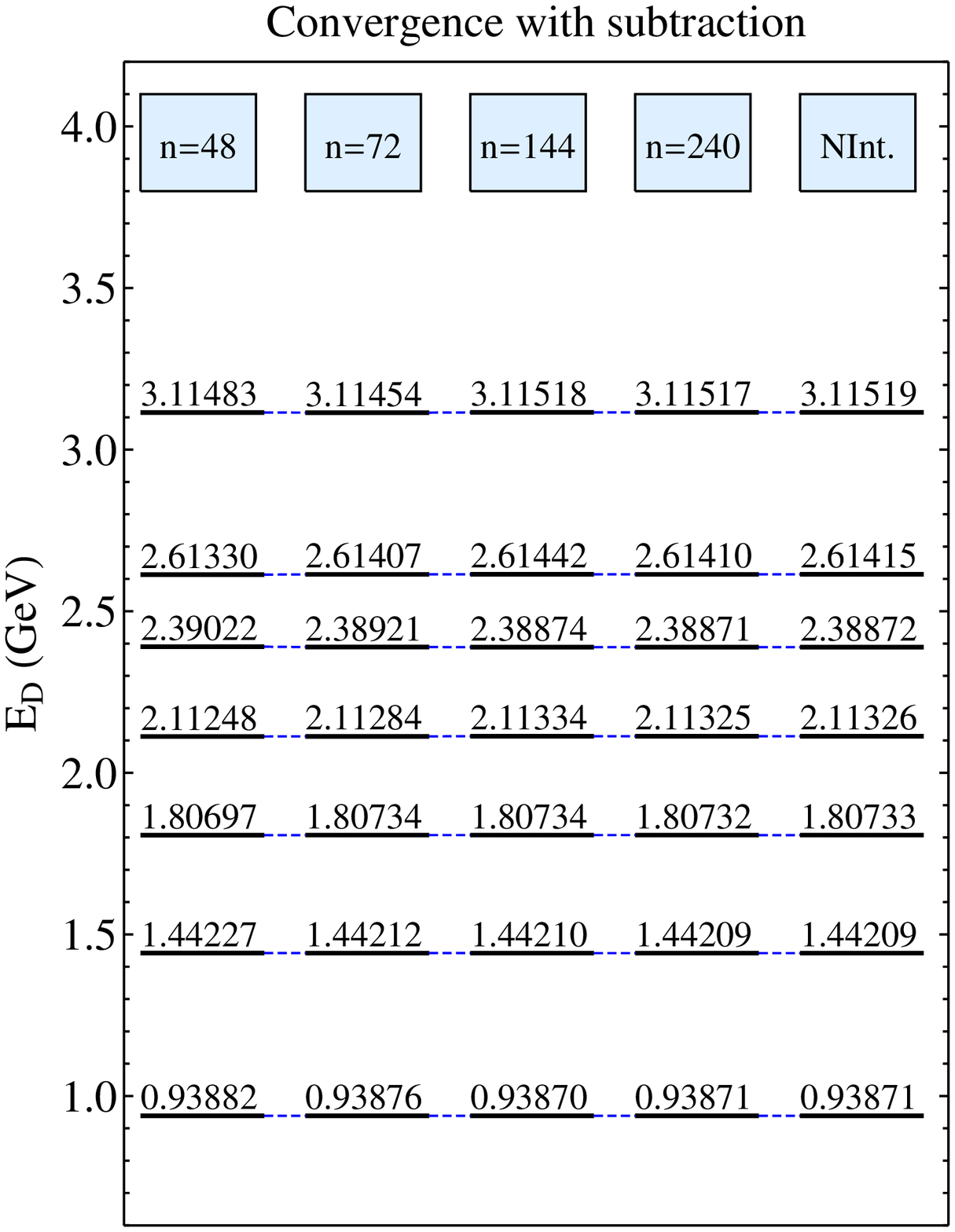}
\centering
\caption{\small Singularity-free 1CSE results.}
\label{with}
\end{minipage}
\end{figure}

\indent In Figs.\ \ref{without} and \ref{with} we study the convergence of the 7 lowest positive energies $E_D\equiv \mu-m_1$ of the 1CSE in the $S$ wave ($\ell=0$) with increasing number of quadrature points $n$.
We consider a mass ratio of $m_1/m_2=5$ (heavy-light meson), $\sigma=0.2$ GeV$^2$ and $y=0$ (pure scalar interaction). The results are computed with 64 B-splines. In Fig.\ \ref{with} the last column "NInt." refers to the results obtained with the adaptive integration routine "NIntegrate" from the software Mathematica 9.0.
\\
\indent We observe a much faster convergence with the singularity-free version of the 1CSE than with the previous unsubtracted version, which is also reflected in a reduction of the computing time to 1/6 to obtain the same accuracy.
%uncomment the following lines to place a figure
\begin{figure}[htb]
\centerline{
\includegraphics[width=7.2cm]{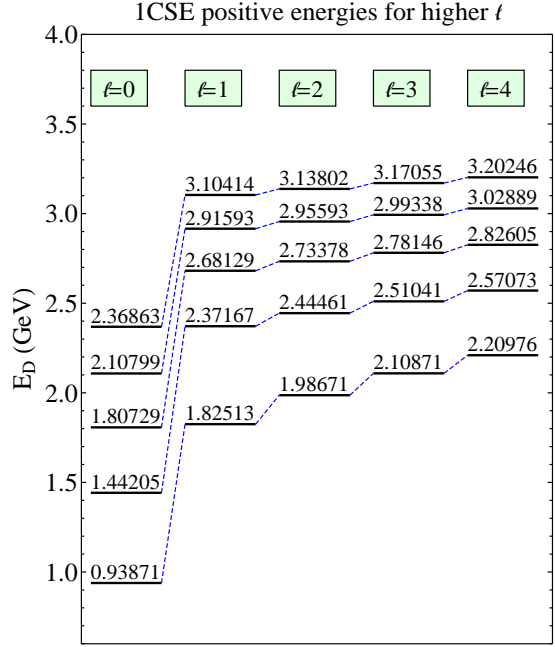}}
\caption{\small Positive energies $E_D$(GeV) for 1CSE  in all partial waves up to $\ell=4$.}
\label{higher}
\label{1CSpartial}
\end{figure}
\\
\indent In Fig.~\ref{higher}, the 5 lowest positive energies $E_D$ of the 1CSE for partial waves up to $\ell=4$ are presented, for the same set of parameters used in Figs.\ \ref{without} and \ref{with}. Even though pseudoscalar particles do not have high orbital angular momentum, these results serve as an important  numerical test, before attempting the study of other structures for the two quark vertex $\Gamma$,  where the high $\ell$ equations become more complex.
 For $\ell>2$ we did not obtain converged results with the unsubtracted version of 1CSE, not even after increasing the number of splines. The subtraction technique fixes this problem because the integrand becomes a smoother function.
\\
\indent In the results of Figs.~\ref{without}, \ref{with}, and \ref{1CSpartial} we neglect retardation and use the simplest replacement
$(p-k)^2\rightarrow - (\mathbf{p}-\mathbf{k})^2$, but we have already performed test calculations with retardation and obtained good convergence.

\indent The main conclusion from this work shows that the numerical method developed in \cite{linear} for the non-relativistic equation is also applicable to the relativistic 1CSE. Future work will generalize it to more complex equations such as the 4CSE, necessary for realistic light meson systems.
\vspace{0.5cm}
\\
\indent This work received financial support from Funda\c c\~ao para a Ci\^encia e a Tecnologia (FCT) under grant Nos. PTDC/FIS/113940/2009, CFTP-FCT (PEst-OE/FIS/U/0777/2013) and POCTI/ISFL/2/275. The research leading to these results has received funding from the European Community's Seventh Framework Programme FP7/2007-2013 under Grant Agreement No. 283286. %The diagrams have been drawn with JaxoDraw \cite{jaxo}.

\end{document}